\documentstyle[prl,aps,epsfig]{revtex}
\def\lsim{\mathrel{\rlap{
\lower4pt\hbox{\hskip-3pt$\sim$}}
    \raise1pt\hbox{$<$}}}     
\def\gsim{\mathrel{\rlap{
\lower4pt\hbox{\hskip-3pt$\sim$}}
    \raise1pt\hbox{$>$}}}     

\begin{document}


 \twocolumn[\hsize\textwidth\columnwidth\hsize  
 \csname @twocolumnfalse\endcsname              

\title{Universal fluctuations in heavy-ion collisions 
in the Fermi energy domain}

\author
{R. Botet$^{\dagger}$, M. P{\l}oszajczak$^{\ddagger}$, 
A. Chbihi$^{\ddagger}$, B. Borderie$^{\S}$, D. Durand$^{\&}$, J.
Frankland$^{\ddagger}$}
 
\address{$^{\dagger}$
Laboratoire de Physique des Solides - CNRS, B\^{a}timent 510, 
Universit\'{e} Paris-Sud, Centre d'Orsay, F-91405 Orsay, France  \\
$^{\ddagger}$
Grand Acc\'{e}l\'{e}rateur National d'Ions Lourds (GANIL), 
CEA/DSM -- CNRS/IN2P3, BP 55027,  F-14076 Caen Cedex, France \\
$^{\S}$ Institut de Physique Nucl\'{e}aire, IN2P3-CNRS, F-91406 Orsay Cedex, 
France \\ 
$^{\&}$ LPC, IN2P3-CNRS, ISMRA et Universit\'{e}, F-14050 Caen Cedex, France}
 
\date{\today}
 
\maketitle
 
\begin{abstract}
We discuss the scaling laws of both the charged fragments multiplicity $n$
fluctuations and the charge of the largest fragment $Z_{max}$
fluctuations for $Xe+Sn$ collisions in the range of bombarding energies 
between $25 \ A\cdot MeV$ and  
$50 \ A\cdot MeV$. We show at $E_{lab} \gsim 32 MeV/A$ the transition in the
fluctuation regime of $Z_{max}$ which is
compatible with the transition from the ordered to disordered phase of
excited nuclear matter. The size (charge) of the largest fragment
is closely related to the order parameter characterizing this process. 
\end{abstract}

\pacs{PACS numbers: 05.70.Jk,24.60.Ky,64.60.Ak,64.60.Fr,64.60.Ht}

 ]  

Theoretical description of the fragment production 
in heavy-ion (HI) collisions depends on whether
the equilibrium has been reached before the system starts fragmenting. 
Possibility of the critical behavior associated with the transition  
from the particle evaporation regime at low
excitation energies to the explosion of the hot 
source at about 5 - 10 MeV/nucleon cannot be excluded.
Unfortunately, this exciting possibility is difficult to study because 
{\it all} standard models and methods of
characterizing different phases and transitions of the nuclear matter 
in HI collisions assume an equilibrium mechanism of the fragment production. 
In this work, we shall apply new methods of the theory of universal
fluctuations of observables in finite systems \cite{order} 
to examine what can be said in a model independent way
about the fragmentation mechanism and the phase changement in HI 
collisions in the Fermi energy domain. Our analysis, which is 
independent of the assumption of the equilibrium in the fragments production
process, uses the data of the INDRA multidetector system for 
$Xe+Sn$ collisions at $25 \ MeV \leq E_{lab}/A \leq 50 \ MeV$ 
\cite{bougault,INDRA,arnaud,salou}. 

Several features of finite systems are important 
if one wants to study either the criticality or 
the distance to the critical point \cite{order}.  
These are : 

(i) The $\Delta$-scaling of the normalized probability 
distribution $P_{<m>}[m]$ of the variable $m$ for different 'system sizes' $<m>$ :
\begin{eqnarray}
<m>^{\Delta}P_{<m>}[m] & = & \Phi (z_{({\Delta})})~,~~ ~~~0 < \Delta \leq
1  \\ \nonumber \\
z_{(\Delta )} & = & (m-m^{*})/<m>^{\Delta} ~ \ ,
\end{eqnarray}
where $<m>$ and $m^{*}$ 
are the average and the most probable values of $m$
respectively, and ${\Phi} (z_{(\Delta )})$ is the positive defined scaling 
function which depends only on a single scaled variable $z_{{(\Delta)}}$.
If the scaling framework holds, the scaling
relation (1) is valid independently of any phenomenological reasons 
for changing $<m>$ \cite{order}. 
The scaling domain in (1) is defined by the asymptotic behaviour of $P_{<m>}[m]$
when $m \rightarrow \infty$ and $<m> \rightarrow \infty$, but $z_{(\Delta)}$ 
has a finite value. The $\Delta-$scaling analysis is very 
robust and can be studied even in small systems if the probability
distributions $P_{<m>}[m]$ are known with a sufficient precision. 
In small systems however, the value of scaling parameter 
$\Delta$ may differ slightly from its asymptotic value \cite{order} ;

(ii) The tail of the scaling function $\Phi(z_{(\Delta)})$  and the anomalous
dimension g.\\
All these features are related to the properties of the 
scaling function which characterizes the finite system. 

If the infinite 
system experiences a second-order phase transition, and if $m$ is the extensive
(scalar) order parameter then \cite{order} :
 
(i) At the critical point, the corresponding finite system exhibits 
the 'first-scaling law' ($\Delta = 1$) 
and the tail of scaling function for large positive $z_{(\Delta)}$ : 
$\Phi(z_{(\Delta)}) \sim \exp(-z_{(\Delta)}^{{\tilde \nu}})$, 
is characterized by a large 
value of the exponent ${\tilde \nu} = 1/(1-g) >2$. 

(ii) The finite system exhibits 
the 'second-scaling law' ($\Delta = 1/2$) in the 
{\it ordered} phase, and the first-scaling law in the {\it disordered} phase. 
In both cases, the tail 
of the scaling function is gaussian (${\tilde \nu}=2$).
Close to the critical point, one may find also the cross-over phenomenon 
from the first-scaling to the second-scaling law by the 
continuous $\Delta$-scaling law, again with the gaussian tail of the 
scaling function.

If the parameter $m$ is not 
singular at the transition, then its probability distribution follows the
second-scaling law with the gaussian tail.

There are two generic families of the fragment production scenarios for which
the second-order phase transition
has been identified. The family of {\it aggregation scenarios}
contains both equilibrium models like 
the Fisher droplet model, the Ising model or
the percolation model, and off-equilibrium models like the 
Smoluchowski model of gelation. In these models,
the average size of the largest cluster $<s_{max}>$ is the order parameter 
\cite{order,aggreg} and the exponent $\tau$ of the power-law cluster-size 
distribution at 
the critical point ($\tau > 2$) : $n(s) \sim s^{-\tau}$,  
is related to  the anomalous dimension as : $g=1/(\tau-1)$.
The second family includes {\it fragmentation scenarios} and contains
\begin{figure}[h]
\centerline{
\psfig{figure=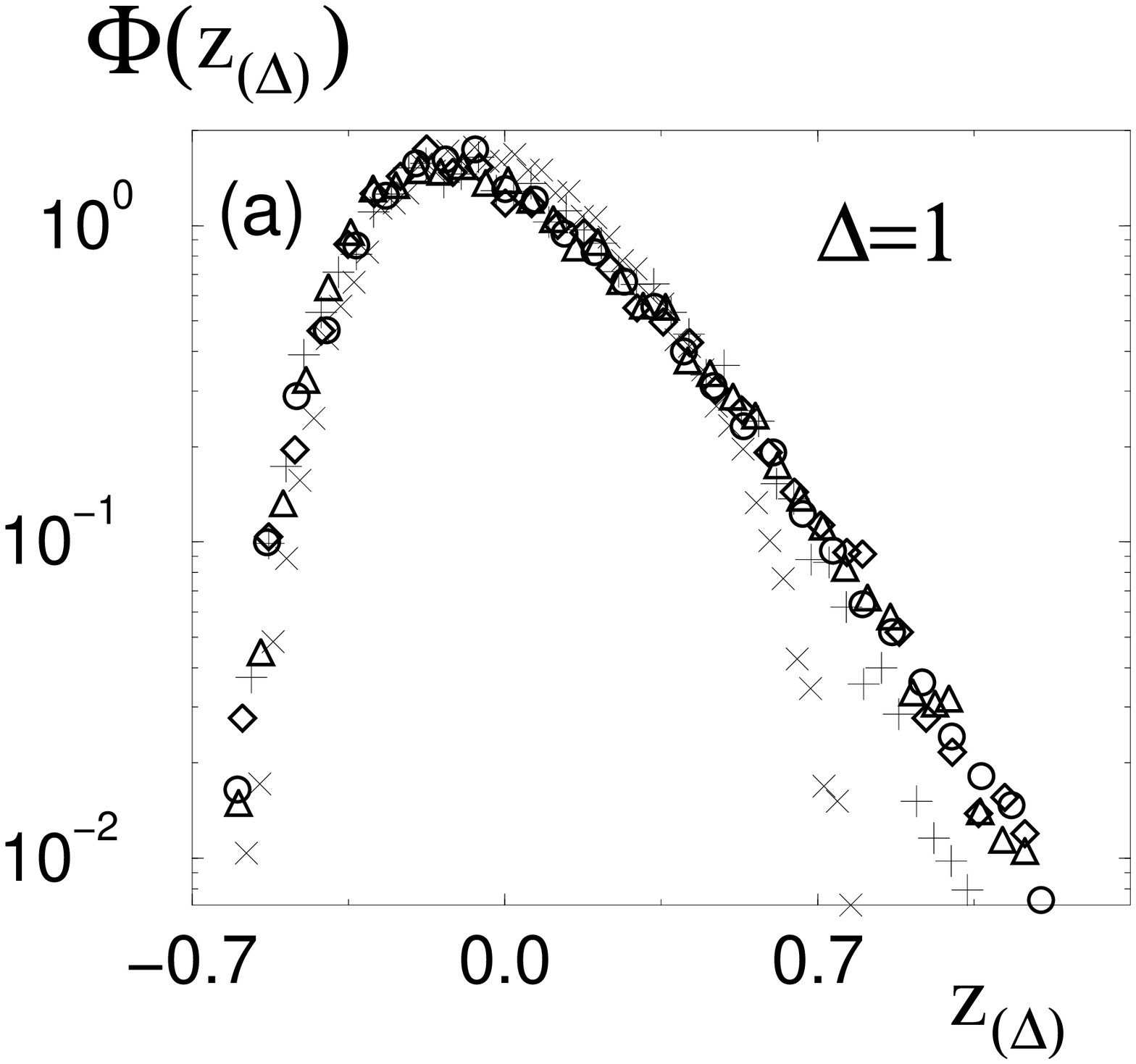,height=5.4cm}
}
\vspace{-4mm}
\centerline{
\psfig{figure=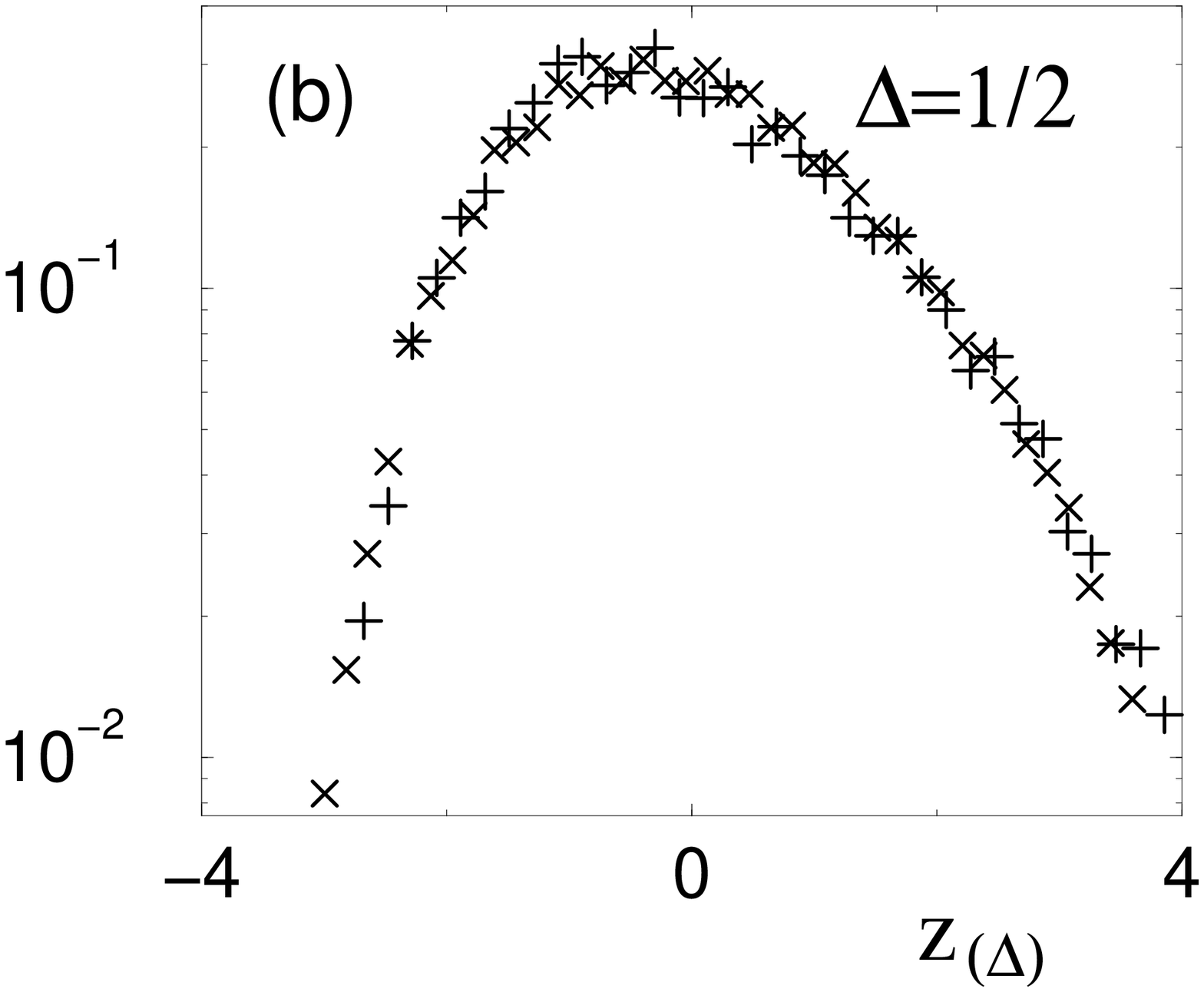,height=5.4cm}
}
\centerline{
\psfig{figure=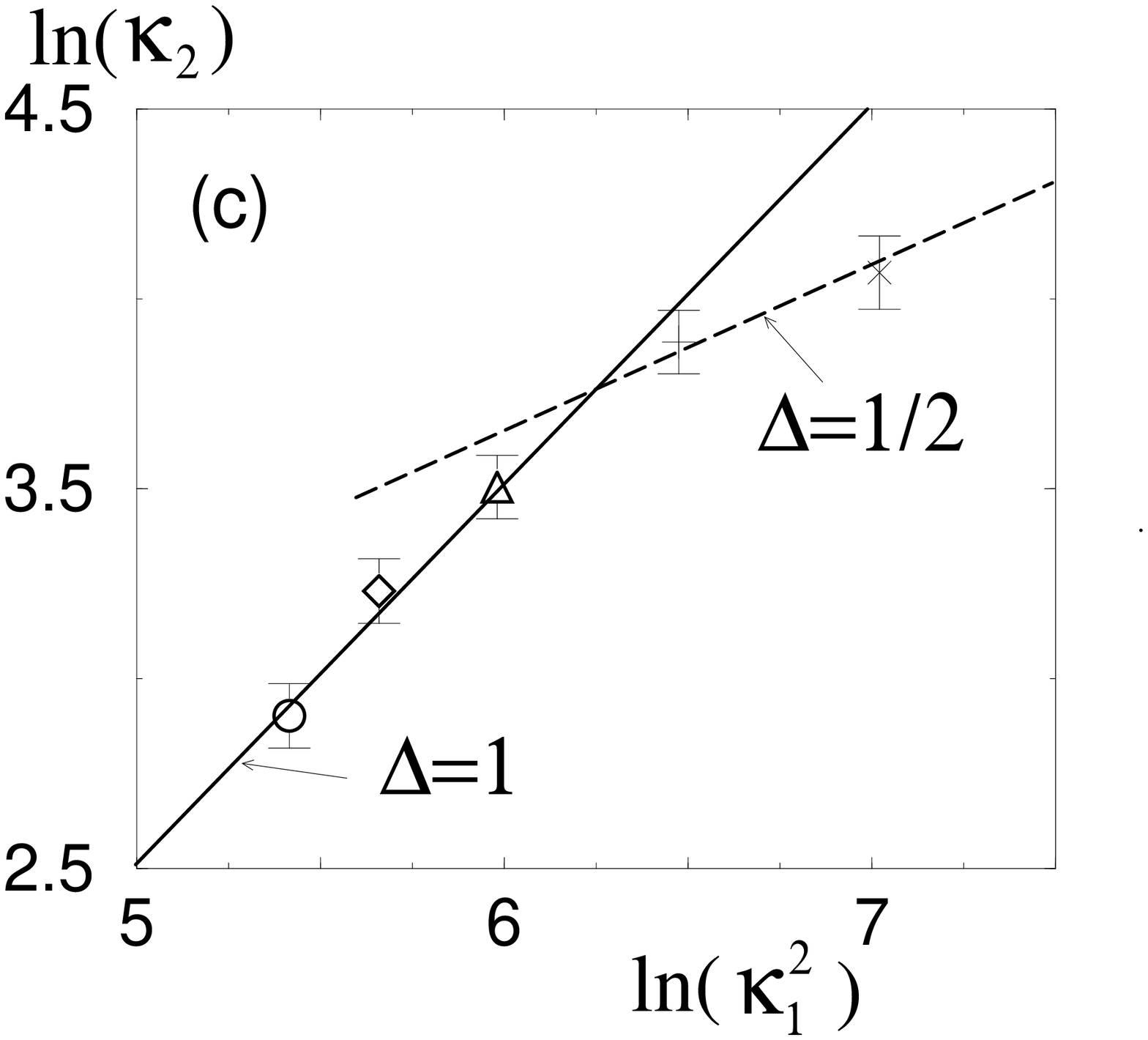,height=5.4cm}
}
\vskip 0.5truecm
\caption{Different characteristics of the largest fragment charge probability 
distributions $P[Z_{max}]$ for central  $Xe + Sn$ collisions.
{\bf (a)} $P[Z_{max}]$ in
the variables of the first-scaling law for
$E_{lab}/A =$ 25 (asterisks), 32 (crosses), 39 (triangles), 45 (diamonds) and 
50 (circles) $MeV$.
{\bf (b)} $P[Z_{max}]$ in
the variables of the second-scaling law for $E_{lab}/A =$ 25 and 32 $MeV$.
{\bf (c)} The normalized second cumulant moment ${\cal K}_2$ 
($\equiv \kappa_2/{\kappa_1}^2$) of $P[Z_{max}]$
is plotted as a function of $\ln <Z_{max}>^2$ ($\equiv \ln \kappa_1^2)$ 
together with the statistical error bars. The lines $\Delta = 1$ 
and $\Delta = 1/2$ are shown to guide the eyes.}
\label{fig1}
\end{figure} 
\noindent
various hybrids of the fragmentation - inactivation - binary model 
\cite{shatt1,shatt2}. In this family of models, the
average fragment multiplicity $<n>$ is the order parameter 
of the critical shattering process and the exponent $\tau$ at the critical
point ($\tau < 2$) is related  to  the anomalous dimension as : $g=\tau-1$.
The order parameters for these two different scenarios are 
not only relevant in the study of phase
changes in HI collisions but they are also measurable.
In this work, we shall
investigate the patterns of the largest fragment charge distribution 
$P[Z_{max}]$  and the charged fragment multiplicity distribution
$P[n]$ for different centrality conditions and bombarding energies 
in the Fermi energy domain, using the
methods of the theory of universal fluctuations of observables in finite
systems \cite{order}. 
 
Fig. 1 shows the $\Delta$-scaling features of $P[Z_{max}]-$ 
distributions for central $Xe+Sn$ collisions at 
$25 \ MeV \leq E_{lab}/A \leq  50 \ MeV$. 
In the experiment a great effort was done to well
identify in charge the different fragments produced and especially the heaviest
ones \cite{ta,pa}. For each collision energy about 20000 events
are taken into account in the present analysis. 
These events are selected with the experimental 
centrality condition : complete events ({\it i.e.}, more than 80\%
of the total charge and momentum is detected) and $\Theta_{flow} \ge\ {\pi}/3$.
The latter quantity is a global observable defined as the angle between the
beam direction and the main emission direction of matter in each event,
which is determined from the energy tensor. It has been shown for the reactions
in the Fermi energy domain that events with small $\Theta_{flow}$ are dominated
by binary dissipative collisions \cite{cugnon,lecolley,INDRA}. For events with
little or no memory of the entrance channel, $\Theta_{flow}$ is isotropically
distributed. The upper part of the figure (Fig. 1a) shows that 
$P[Z_{max}]-$distributions for $E_{lab}/A = 39, 45, 50 \ MeV$ can be 
compressed into a single curve in the scaling variables of the first-scaling. 
The distributions for 25 and 32 $MeV$, which show strong deviations with 
respect to this scaling curve both near the maximum and in the tail for 
large $z_{(\Delta)}$, can be compressed into another single curve
in the variables of the second-scaling, as shown in Fig. 1b. It should be
stressed that we do not optimize the value of $\Delta$ because of the
experimental (number of events) and theoretical 
(smallness of the system) limits, but we study whether the data is
consistent with one of the two limits : $\Delta = 1/2$ and 1, which 
have a particular significance in the scaling theory of phase-transitions. 

More straightforward global measure of scaling features
is provided by the  cumulant moments :
$\kappa_1=<m>, \kappa_2=<m^2>-<m>^2, \kappa_3=<m^3>-3<m^2><m>+2<m>^3$, 
{\it etc.} In case of the $\Delta$-scaling, normalized cumulant moments :
${\cal K}_q^{(\Delta)} \equiv {\kappa_q}/{(\kappa_1)^{q \Delta }} \ ,$
are independent of the system size $<m>$
\cite{order}. Log of the normalized cumulant moment 
${\cal K} \equiv {\cal K}_2^{(\Delta = 1)}$ of 
$P[Z_{max}]$ is plotted in Fig. 1c versus log of $<Z_{max}>^2$, 
{\it i.e.} versus log of $(\kappa_1)^2$.
The data for different $<Z_{max}>$, {\it i.e.} for 
different bombarding energies, 
should lie on a straight line if the $\Delta$-scaling holds. 
The slope of this line gives the value of $\Delta$. 
It is seen that the higher energy branch ($E_{lab}/A = 39, 45, 50 \ MeV$) 
follows the line $\Delta=1$ (the solid line), in agreement 
with Fig. 1a. The point for $E_{lab}/A = 25 \ MeV$ is clearly off this line. 
The point for $E_{lab}/A = 32 \ MeV$ aligns 
along the line $\Delta=1/2$ (the dashed line) 
passing through the point for $E_{lab}/A = 25 \ MeV$, but is also close to the 
line $\Delta=1$. In this 'near-crossing' case, 
one has to investigate higher order moments, viz the whole probability 
distribution. Fortunately, Figs. 1a, 1b show clearly that the  
collisions at 32 $MeV/A$ belong to the branch $\Delta = 1/2$. 
\begin{figure}[h]
\centerline{
\psfig{figure=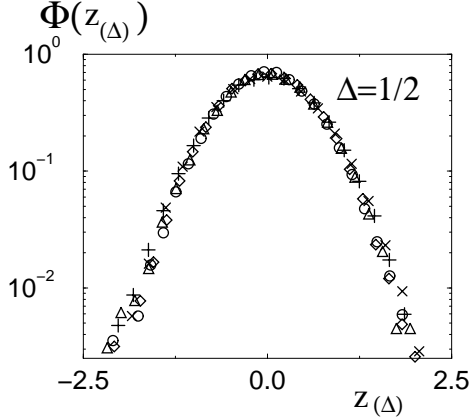,height=5.4cm}
}
\vskip 0.5truecm
\caption{The multiplicity distributions of charged fragments 
for central $Xe + Sn$ collisions are plotted in  the variables  
of the second-scaling law. Notation of data at
different collision energies is the same as in Fig. 1.}
\label{fig2}
\end{figure} 

The pattern of charged fragment multiplicity distributions $P[n]$
does not show any significant evolution with the bombarding
energy (Fig. 2), and the data is perfectly compressible
in the scaling variables of the second-scaling,
{\it i.e.} the multiplicity fluctuations are small ($\kappa_q \sim
(\kappa_1)^{q/2}$) in the whole studied range of
bombarding energies. 
The scaling features of experimental $P[Z_{max}]-$ and $P[n]-$ probability 
distributions in Figs. 1 and 2 are complementary and allow to affirm 
that the fragment production in central HI reactions 
in the Fermi energy domain follows the aggregation scenario 
and exhibits the transition at $E_{lab}/A \gsim 32 \ MeV$ between the 
two phases of excited nuclear matter
with distinctly different patterns of $Z_{max}-$fluctuations. 

Assuming that the change of the fragmentation regime is controlled 
by the source excitation energy, one may ask to which extent 
similar scaling features and transitions can be seen 
in more peripheral collisions. For
that purpose, we have investigated a sample
of events selected with the experimental condition :
complete events and 
$\Theta_{flow} \ge\ {\pi}/18$. This sample contains mostly 
events of semi-central binary collisions which keep some memory of the entrance
channel as manifested by the presence of two contributions in the velocity
distribution of $Z_{max}$ 
corresponding to the quasi-projectile and quasi-target \cite{salou}. 
Fig. 3a shows that $Z_{max}-$distributions for 
$E_{lab}/A =$ 45 and 50 $MeV$ can be compressed into a single curve in the
variables of the first-scaling. The distributions for 
\begin{figure}[h]
\centerline{
\psfig{figure=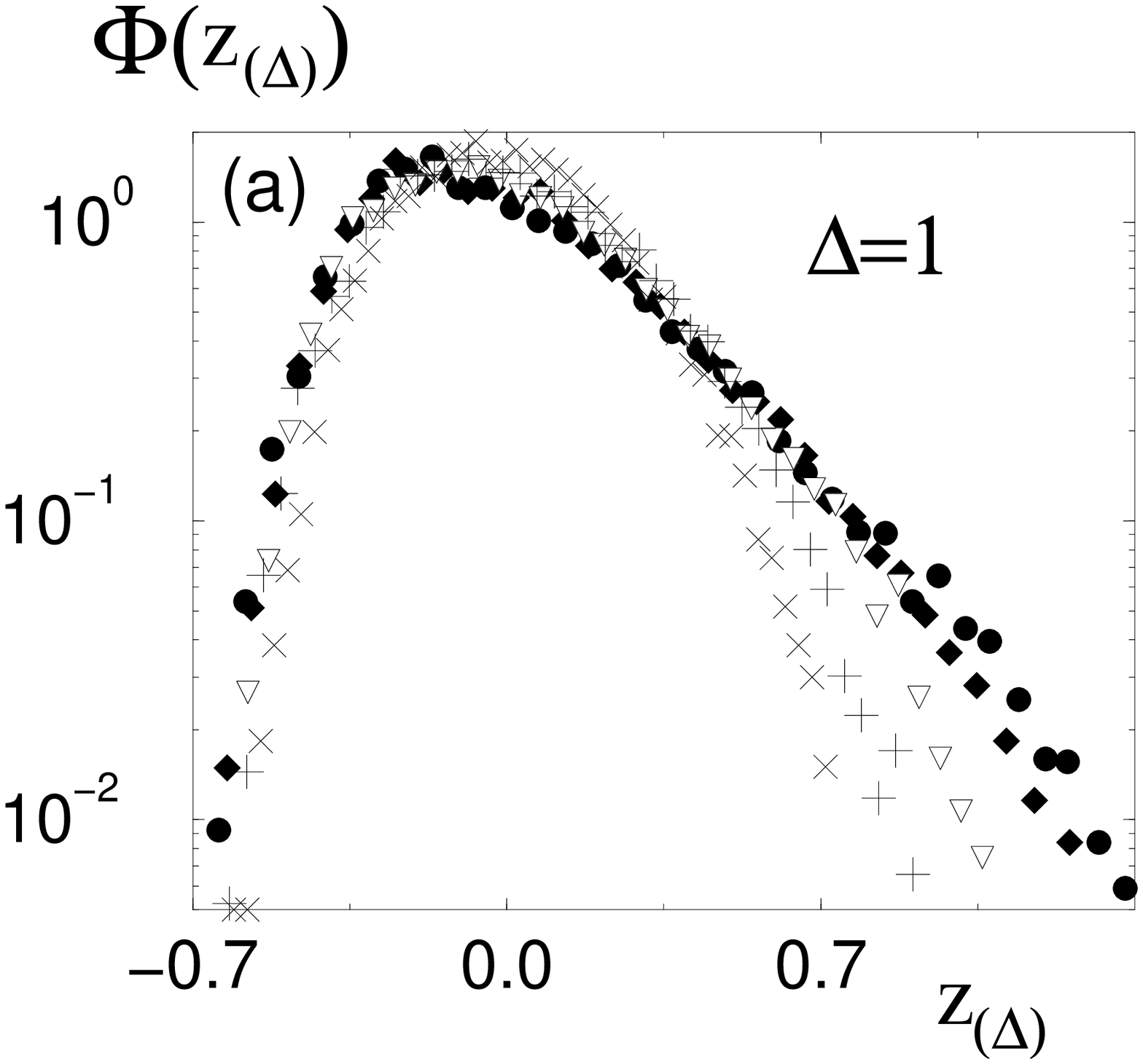,height=5.5cm}
}
\vspace{-4mm}
\centerline{
\psfig{figure=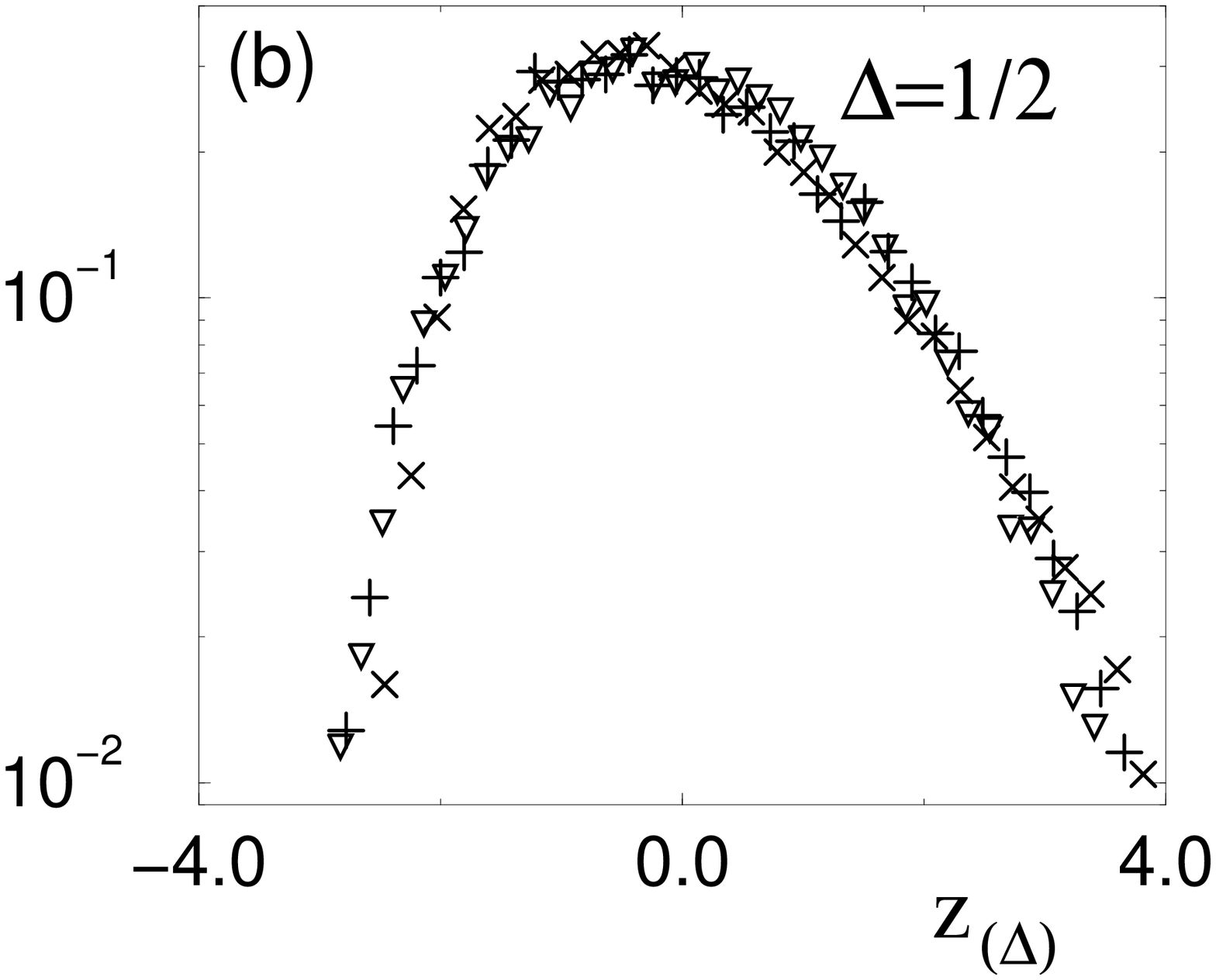,height=5.5cm}
}
\vspace{-4mm}
\centerline{
\psfig{figure=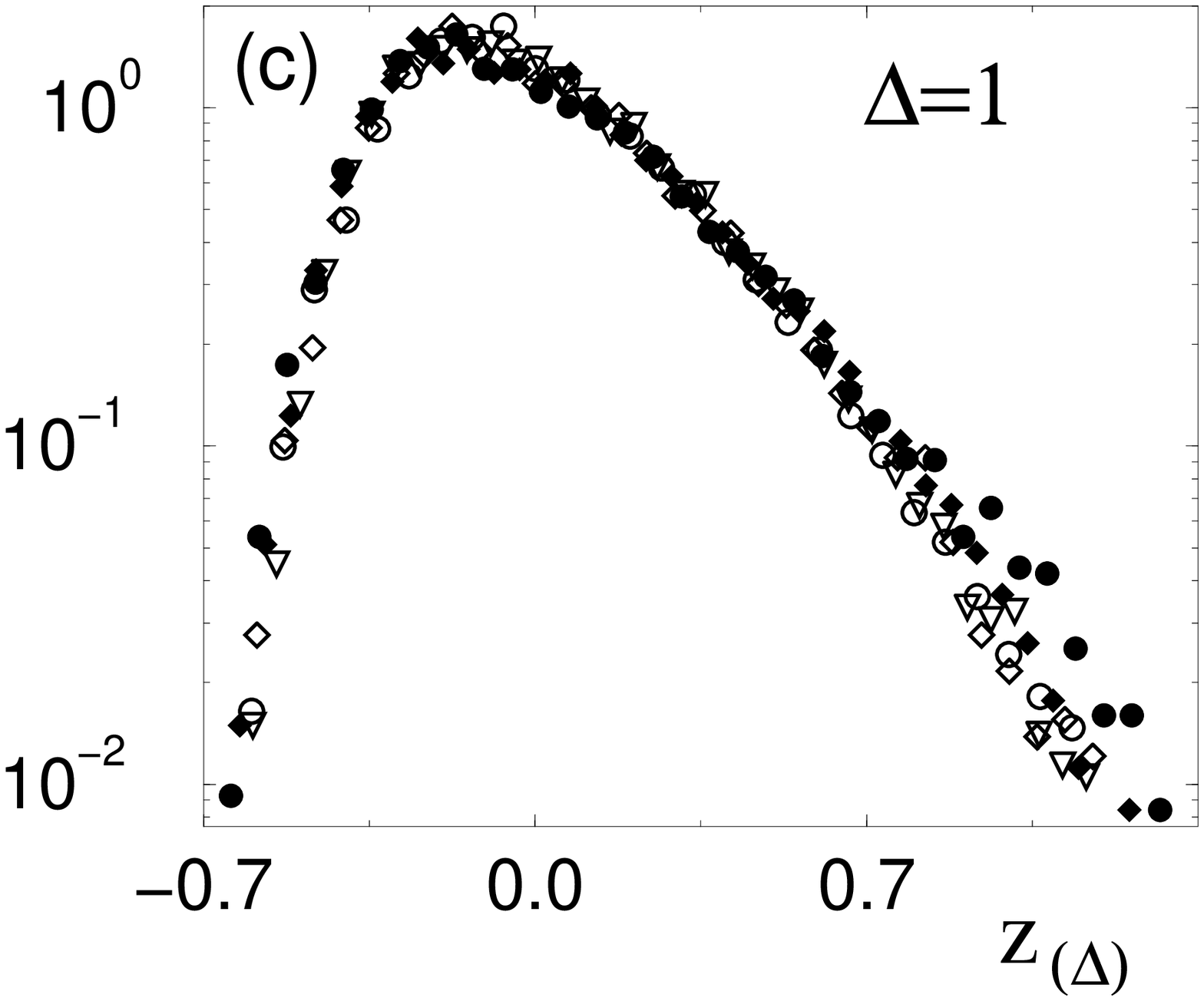,height=5.5cm
}
}
\vskip 0.5truecm
\caption{Different characteristics of the largest fragment charge  
distributions $P[Z_{max}]$ for $Xe + Sn$ collisions.
{\bf (a)} $P[Z_{max}]$ for semi-central collisions in
the variables of the first-scaling law for
$E_{lab}/A =$ 25 (asterisks), 32 (crosses), 39 (triangles), 45 (full diamonds)
and 50 (full circles) $MeV$.
{\bf (b)} $P[Z_{max}]$ for semi-central collisions at 
$E_{lab}/A =$ 25, 32 and 39 $MeV$ in
the variables of the second-scaling law.
{\bf (c)} $P[Z_{max}]$ for semi-central collisions at 
$E_{lab}/A =$ 45 and 50 $MeV$ and central collisions at 
$E_{lab}/A =$ 39, 45 and 50 $MeV$ in
the variables of the first-scaling law. The notation for central
collisions is the same as in Fig. 1.}
\label{fig3}
\end{figure}
\noindent
25, 32 and 39 $MeV$ show
significant deviations with respect to this scaling curve, both at the
maximum and in the large-$z_{(\Delta)}$ tail. These three distributions are
shown in Fig. 3b in the variables of the second-scaling.
$P[Z_{max}]-$distributions for $E_{lab}/A =$ 25 and 32 $MeV$ can be perfectly
compressed into a single curve. The data for $E_{lab}/A =$ 39 $MeV$ is close to
this curve but, nevertheless, shows some deviations for
$z_{(1/2)} \simeq 1$. This data, which seems to be intermediate between the scaling
limits $\Delta = 1/2$ and $\Delta = 1$, could indicate a continuous 
change of $\Delta$ in the transition region. 
There are presently not enough data to verify this possibility. 
We note however, 
that an optimal data compression for $E_{lab}/A =$ 32
and 39 $MeV$ would indicate $\Delta \simeq 0.6$ for the latter one. 

The scaling pattern of $P[Z_{max}]$ in semi-peripheral collisions
follows a  similar general evolution with the bombarding energy 
(the excitation energy) as seen in central collisions, 
except that the branch $\Delta = 1$ in semi-central collisions 
starts at higher energies than in central collisions. Fig. 3c presents 
all $Z_{max}-$distributions which 
in semi-central and in central collisions belong to the respective 
first-scaling branches. It is non-trivial that these probability 
distributions, which correspond to
different selection criteria, collapse approximately into  
the unique scaling curve. This example should encourage further 
investigation of the feasibility of semi-central events in this kind of studies.
One should also note that the scaling functions in the 
branch $\Delta = 1/2$ are essentially identical in semi-central
and central collisions (compare Figs. 1b and 3b). 
 
The characteristic feature of critical behavior
is the anomalous tail of the scaling function with ${\tilde \nu}>2$.
Using all $P[Z_{max}]$ shown in Fig. 3c, we have fitted 
the scaling function $\Phi(z_{(1)})$
for $z_{(1)}>0$ by : 
$a \exp (-b (z_{(1)}-z_0)^{\tilde \nu})$, where $z_0$ is the 
estimate of the most probable value of the
distribution and $a, b, {\tilde \nu}$ are the fitting parameters. We have found : 
${\tilde \nu} = 1.6 \pm 0.4$, what is incompatible with typical 
values $(3.5 \lsim {\tilde \nu} \lsim 6)$ in the critical region for
aggregation scenarios \cite{order}. The same procedure 
for central and semi-central collisions
in the second-scaling phase yields : 
${\tilde \nu} = 1.8 \pm 0.4$. Hence, the existing data do not show the 
critical bahavior in the transition region from ordered to disordered phase.

In conclusion, we have applied the theory of
universal fluctuations in the finite systems \cite{order} 
to the symmetric HI reactions in the Fermi energy domain.
This theory provides rigorous methods to characterize critical and off-critical
behaviors both in equilibrium and off-equilibrium finite systems. 
This is an important novel aspect of the present analysis.
Convincing scaling behavior of the $Z_{max}-$ and $n-$distributions have been
found in central collisions. The present analysis shows that the fragment
production in central HI collisions at around the Fermi energy 
is governed by the aggregation scenario with $<Z_{max}>$ as the order
parameter. The changement of the regime of $Z_{max}-$fluctuations 
from the second-scaling at low energies to
the first-scaling at higher energies, with the gaussian tail of the scaling
function in both regimes, is compatible with the 
transition from the ordered phase to the disordered phase. 
The scaling curves determined from central and semi-central collisions are the
same but the bombarding energy corresponding to change in the scaling behavior
is higher for more peripheral collisions, in according with the expected 
stronger contribution of non-equilibrium effects in these collisions. 
The critical region, if exists in the nuclear fragmentation process,
should be searched for in the narrow window of bombarding
energies close to : $E_{lab}/A \sim 32 \ MeV$. Its signature would be the
anomalous tail ${\tilde \nu}>2$ of $P[Z_{max}]$, in case of the second-order phase
transition, or the double-hump shape of $P[Z_{max}]-$distribution 
for the first-order phase transition \cite{order}. 
Future studies of symmetric HI reactions,
performed in small steps of bombarding energies,  
will hopefully allow to distinguish between the cross-over phenomenon 
and the other two scenarios which invoke existence of the phase-transition.

\vskip 0.2 truecm
{\bf Acknowledgements}\\
\noindent
We are grateful to the Collaboration INDRA for providing us the experimental 
data. We thank J.-L. Charvet and J. Colin for stimulating discussions.

\end{document}